# Microwave Nonlinearity and Photoresponse of Superconducting Resonators with Columnar Defect Micro-Channels

S.K. Remillard[1,4], D. Kirkendall[1], G. Ghigo[2,3], R. Gerbaldo[2,3], L. Gozzelino[2,3], F. Laviano[2,3], Z. Yang[4], N.A. Mendelsohn[4], B.G. Ghamsari[4,6], B. Friedman[4], P. Jung[5], S.M. Anlage[4]

[1]Physics Department, Hope College, Holland, MI 49423, USA
[2]Politecnico di Torino, Dept. of Applied Science and Technology, c.so Duca degli Abruzzi 24,10129 Torino, Italy
[3]Istituto Nazionale di Fisica Nucleare, Sezione di Torino, via P. Giuria 1, 10125 Torino, Italy
[4]Center for Nanophysics and Advanced Materials, University of Maryland, College Park, MD 20742, USA
[5]Physikalisches Institut, Karlsruhe Institute of Technology, Wolfgang-Gaede-Str. 1, D-76131 Karlsruhe, Germany
[6]School of Electrical Engineering and Computer Science, University of Ottawa, Ottawa, ON K1N 6N5, Canada.

## Abstract

Micro-channels of nanosized columnar tracks were planted by heavy-ion irradiation into superconducting microwave microstrip resonators that were patterned from $YBa_2Cu_3O_{7-x}$ thin films on $LaAlO_3$ substrates. Three different ion fluences were used, producing different column densities, with each fluence having a successively greater impact on the microwave nonlinearity of the device, as compared to a control sample. Photoresponse images made with a 638 nm rastered laser beam revealed that the channel is a location of enhanced photoresponse and a hot spot for the generation of intermodulation distortion. The microwave photoresponse technique was also advanced in this work by investigating the role of coupling strength on the distribution of photoresponse between inductive and resistive components.

## 1. Introduction

Superconducting microwave transmission line resonators exhibit microwave nonlinearity which could be harnessed for low noise detection[1], restrained for passive microwave filters[2], and rendered purely inductive for quantum read-out devices[3]. Columnar defects in cuprate superconductors caused by heavy ion irradiation have been found to reduce nonlinearity by increasing the critical current density for the onset of nonlinearity[4]. Conforming the heavy ion radiation to the current distribution (e.g. using a larger dose where the current in the device is higher) even further suppresses the inductive nonlinearity, and the linear behavior persists to even higher current[5]. Although columnar defects serve as pinning sites that arrest the motion of magnetic vortices[6], a high density of such defects suppresses superconductivity which then manifests in a reduced local critical temperature $T_C$. When the heavy-ion beam is micro-collimated, confined microchannel regions can be created[7] where the local order parameter is reduced, opening up potential application to magnetic field sensors[8] and THz detectors[9]. These high defect density channels exhibit an enhanced nonlinear Meissner effect (NLME) and microwave devices incorporating these microchannels are highly nonlinear.

In this work, a patterned superconducting transmission line resonator implanted with a microchannel was used in order to investigate the nonlinear electrodynamics of the microchannel. Two effects of nonlinearity are (1) a microwave current dependent Q factor and resonant frequency [10], and (2) frequency effects of the current leading to harmonic and intermodulation distortion (IMD) [11]. These nonlinear effects have informed our understanding of superconducting electrodynamics. But because these are global measurements which take an average across the sample, weighted by the current distribution, they give limited information when examining electrodynamics on a micron scale, especially in the vicinity of a microchannel.

In order to address the limitation of global measurement, spatially resolved microwave measurements of superconductors gained interest beginning in the 1990s. Using confocal resonators, spatial maps of the surface impedance [12] have served as a useful diagnostic for developing large area high quality epitaxial superconducting thin films [13]. Sheet resistance with resolution better than 100 μm was measured using near-field microwave microscopy [14,15]. Also, near field microwave microscopy using magnetic loop probes [16,17] or a high resolution magnetic write-head [18] and scanning three-tone excitation [19] have been used to map harmonic generation and IMD, respectively, which increase as the nonlinearity scaling current density $J_{NL}$ decreases, a behavior understood to occur around microscopic material defects.

The thermal, or bolometric, effect of laser light on superconducting thin films [20] initially motivated the use of photoresponse (PR) as a local probe of superconducting resonators by scanning across the transmission line [21]. More recently, the PR of the microwave conductivity was used for dimensional imaging of the local microwave current density in a superconductor as well as two dimensional images of the local IMD [22]. The anisotropy in the nonlinear Meissner effect in *d*-wave cuprate superconductors was directly observed with such imaging [23].

In this paper, microchannels of columnar defects are introduced into YBCO microstrip microwave resonators by heavy ion irradiation. The microwave nonlinearity of the microchannels was examined first globally by Q measurement, but then locally by scanning IMD and PR. The IMD scans reveal the location of IMD generation; whereas the PR scans indicate with very high resolution whether this nonlinearity is inductive or resistive. We will see close correlation between the IMD and the PR, and evidence from PR that higher ion beam fluence results in elevated resistive nonlinearity in the channel, and that this higher resistive nonlinearity is a local source of the enhanced nonlinearity.

| Sample | Fluence | f (MHz) | $Q_u$ |
|--------|---------|---------|-------|
| 1 | $5\times10^{11}$ | 832.7 | 14,000 |
| 2 | $2.5\times10^{11}$ | 838.3 | 15,000 |
| 3 | 0 | 838.5 | 12,500 |
| 4 | $7.5\times10^{11}$ | NA | NA |

**Table 1.** Summary of the four samples used in this study indicating the Au ion beam fluence, the fundamental resonant frequency and its unloaded Q at 77 Kelvin and low microwave power.

## 2. Samples

The samples were originally fabricated for use in commercial microwave filters for wireless base stations. Thin films of $YBa_2Cu_3O_{7-x}$ (YBCO) sputtered onto both sides of a $LaAlO_3$ substrate were patterned into the resonator in Figure 1 with 250 μm linewidths. Photoresist was spin coated onto the 400 nm thick YBCO film, baked, exposed under a mask to UV light, and milled for approximately 40 minutes with a 70 mA Ar ion beam. After patterning, the wafer was annealed in $O_2$ at $500^oC$ for one hour. With one side unpatterned, the device was operated in a microstrip geometry. The unpatterned ground-plane side was coated with gold so that the diced wafer could be indium soldered to a gold plated titanium carrier.

A beam of 250 MeV $Au^{197}$ ions was used to modify the transmission line. By means of a stainless steel micro-collimator, a 55 μm wide channel was created with a uniform distribution of ion-induced defects, which are mainly columns of amorphized material with nanometric cross-section forming along the ion track. The channel material shows a slight topographic elevation from the rest of the film[24] due to strain induced by implantation of the ions into the substrate to a depth of about 13 μm below the film surface. Moreover, the columnar defects impose non-superconducting regions in the film, thus suppressing the carrier density in the channel. The channel, seen in Figure 1, was located midway between the two ends of the transmission line.

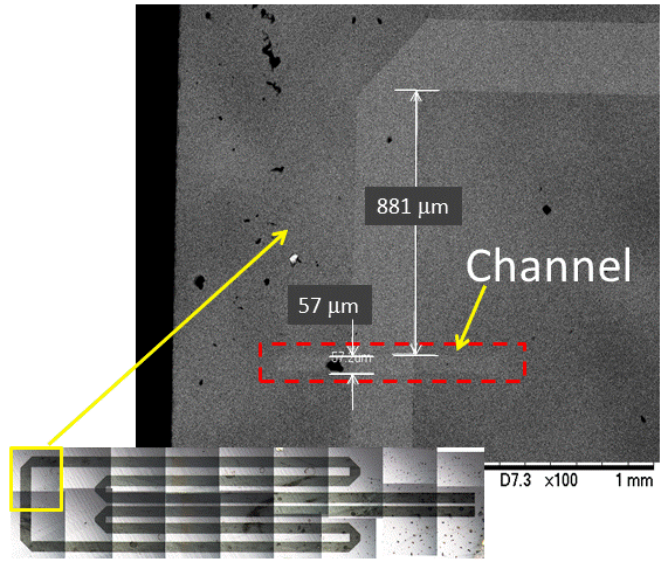

**Figure 1.** SEM micrograph of the 250 μm linewidth resonator used in this experiment. The close-up view was taken with a polarized light microscope and the channel formed by the ion beam is visible. The light microscope photo at the bottom was pieced together from several images with dark regions showing the YBCO film while light regions show the exposed substrate. The entire structure from left to right is 4.5 mm long.

Three identical resonators, summarized in Table 1, were used in this study, each one exposed to a different Au ion fluence. **Sample 1** was irradiated with a fluence of $5\times10^{11}$ cm$^{-2}$, corresponding to a dose equivalent field (the magnetic field that is required to fill each track with one flux quantum) of $B_\phi$=10T; **Sample 2** was irradiated with a fluence of $2.5\times10^{11}$ cm$^{-2}$ ($B_\phi$=5T); **Sample 3** (the control sample) was not exposed. A fourth sample (**Sample 4**) was irradiated with a fluence of $7.5\times10^{11}$ cm$^{-2}$ ($B_\phi$=15T), which rendered the channel non-superconducting, completely suppressing the fundamental resonance, though not the next resonance at 2.0 GHz.

The fluences were high enough to produce a local decrease of the bulk critical current *Jc* as opposed to low-fluence experiments where columnar defects are used to improve $J_C$[25]. At these fluences the critical temperature $T_C$ is also reduced in the channel region, mainly due to secondary electrons affecting the regions around the amorphous columns and to irradiation induced strain in the substrate[24]. Furthermore, the presence of the amorphous columnar defects reduces the volume of the film that actually supports superconductivity, resulting in local suppression of the carrier density in the channel which influences the NLME.

Channels were previously shown to impact the fundamental resonance (first mode) of the line, but not the second mode [26]. Although superconductors typically respond nonlinearly to induced microwave current, the unloaded quality factor, $Q_u$, of the fundamental mode of Sample 1 at 835 MHz, shown in Figure 2, begins to respond at a power two orders of magnitude lower than the Q of the 2$^{nd}$ resonant mode at 2.0 GHz, as gauged by a 25% increase in $1/Q_u$. The fundamental, with peak current at the channel, is more dissipative at elevated RF current ($\propto \sqrt{P_d}$) than the second mode is, which has no current at the channel. An even more striking comparison is made between the fundamental of Sample 1 and the fundamental of the control Sample 3, which requires four orders of magnitude more power than Sample 1 to induce a 25% increase in $1/Q_u$ of its fundamental. $Q_u$ is a macroscopic quantity sensitive only to a weighted average surface resistance of the entire sample, whereas $T_C$ is sensitive to the weakest portion. However, given that the fundamental has peak current at the channel and the 2$^{nd}$ mode has a current null at the channel, the contribution of the channel to dissipation is unmistakable. As microwave current crosses the channel, it encounters a small region of both elevated dissipation and enhanced nonlinearity.

## 3. Intermodulation Distortion (IMD)

IMD is more sensitive than Q to nonlinearity since IMD is detected at microwave currents several orders of magnitude below that where the Q begins to change. IMD is usually a macroscopic measurement, revealing only the average nonlinearity of the device under test. Two methods have been used in this work for scanning the local IMD. In the first method a raster probe introduces non-resonant tones which generate IMD only at the probe's location and allows for the measurement of the current at the IMD frequency[19]. Another method, described in the next section, uses a laser scanning microscope (LSM) to provide a high resolution map of 3rd order IMD[22].

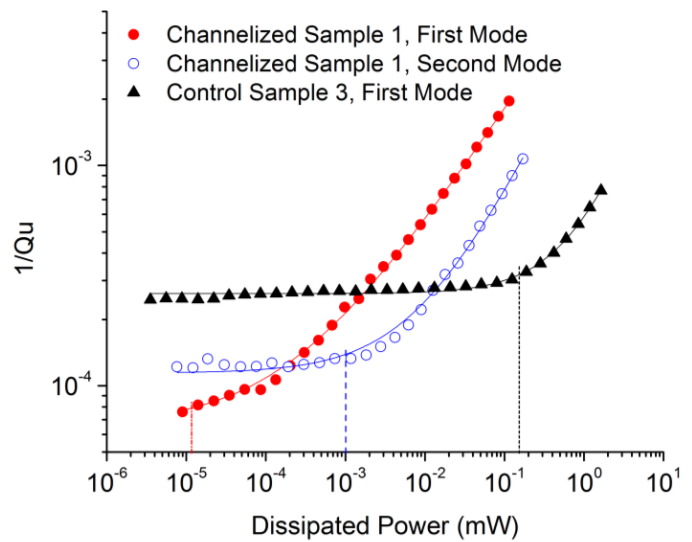

**Figure 2.** The inverse $Q_u$, proportional to the average $R_S$, for the first two resonant modes (835 MHz and 2,000 MHz) of Sample 1 at 77K are shown and the first resonant mode of the control Sample 3. The dashed lines show the dissipated power ($\propto H_{RF}^2$) where $1/Q_u$ has risen to 25% above the residual level given by the power law fit.

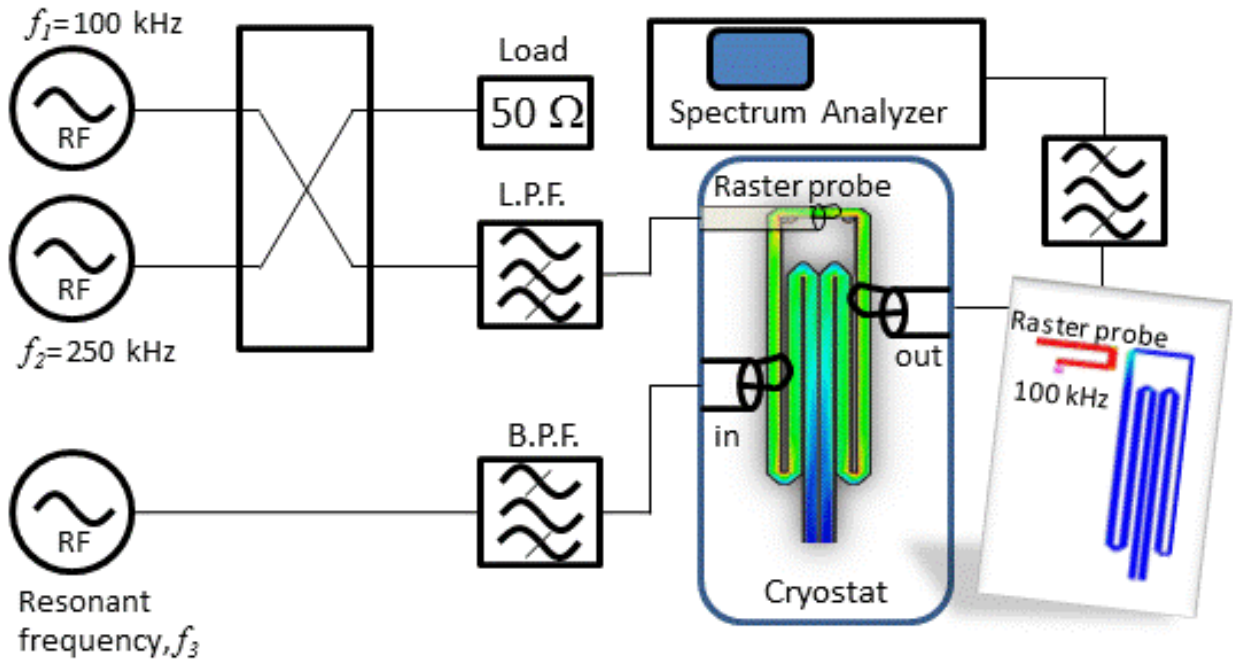

**Figure 3.** (colour on-line) The Three-tone IMD measurement involves three loop probes. The raster probe excites current which remains localized at the probe, as shown by the method-of-moments IE3D simulation of current density in the inset on the right. The IMD output is measured with the fixed probe labelled "out". Also shown is a simulation of the resonance mode current distribution, also done using IE3D.

In order to locally measure IMD, a signal at $f_1$ far from resonance is introduced through a small probe which is scanned over the sample. Because $f_1$ does not excite resonance, near field current is induced only locally and the $f_1$ signal does not propagate. The inset in Figure 3 shows a method-of-moments simulation of the current induced in a transmission line resonator by a probe carrying out-of-band power at 100 kHz. A signal $f_3$ at the resonant frequency is introduced by a stationary input probe (labelled "in" in Figure 3). The current in the resonant mode mixes with the out-of-band local probing current generating IMD at $2f_3 \pm f_1$ only where there is probing current, e.g. within proximity of the raster probe. The 3rd order IMD at $2f_3 \pm f_1$ is not in the resonant band and therefore does not propagate to the output probe. This inconvenience is remedied by scanning with two out-of-band tones $f_1$ and $f_2$ closely spaced such that $f_2-f_1 \ll \delta f$ where $\delta f$ is the 3 dB bandwidth of the resonator. Now, the third order IMD occurs at $f_3-(f_2-f_1)$, which is in-band of the resonator making the locally generated IMD a source of mode excitation.

Electromagnetic field simulation, in this work using HFSS (Ansys, Inc., Canonsburg, PA, USA), is at the heart of converting the detected IMD power into the surface current density that generates the IMD, and is described in Reference [19]. Briefly, field simulation at resonance provides a current density profile $K(\ell)$ along the transmission line, such as shown in Figure 3, computed for the convenient case of 1 Watt of input power. $\ell$ is a linear coordinate along the resonant line. The dissipated power, which is easily computed from the loaded Q, and coupling coefficients[19], versus measured output power is converted into a function $K(P_{out})$, where $P_{out}$ is measured with the spectrum analyser at the resonance peak. The surface current density that produces the measured IMD power is then determined by inserting the IMD output $P_{IMD,}$ instead of the carrier output $P_{out}$, into the function $K(P_{IMD})$.

When intrinsic effects such as the NLME and flux penetration dominate the nonlinearity, third order IMD, as well as third order harmonic distortion, will have a slope versus input power (in dBm) of 3 [27,28]. References [27] and [28] examined this carefully in terms of induced currents and circulating power because (i) RF coupling into the nonlinear superconducting device changes with power, and (ii) $J_{RF}^2$ is not exactly proportional to input power. This is true above a threshold input power where both the Q and the input/output coupling change. Extrinsic mechanisms, such as weak links, result in a smaller slope[28].

In the case of three-tone IMD, only the one in-band signal power is swept and the expected slope for intrinsic nonlinearity is 1. With the probe over the channel in Figure 4a, the low power third order IMD has a slope of 0.83±0.02 determined by fitting a power law model to the data and its uncertainties in Origin 8.0 (OriginLab Corp, Northampton, MA, USA). Under the same conditions, the control sample has a somewhat larger slope of 0.94±0.03, suggesting that there is likely more extrinsic nonlinearity in the channel than in the as-grown film. At higher power, the IMD of all samples reaches a peak and then drops as Q begins to decrease. This peak occurs three orders of magnitude higher in input power for Sample 3 than for Sample 1. IMD drops more dramatically at high power whenever the probe is over a hot spot in IMD[29]. Above the power where Q begins to degrade, the channel's IMD decreases with a slope of -1.8±0.1 compared to the control sample which has a smaller slope of -1.0±0.1 and is similar to the behavior seen in local third harmonic measurements of bulk niobium[30]. A saturation and possible decline is also seen in harmonic balance simulation using the intrinsic nonlinearity of the NLME with a quadratic current dependence of resistance and inductance [31,32] as well as in other forms of behavioral modelling[33].

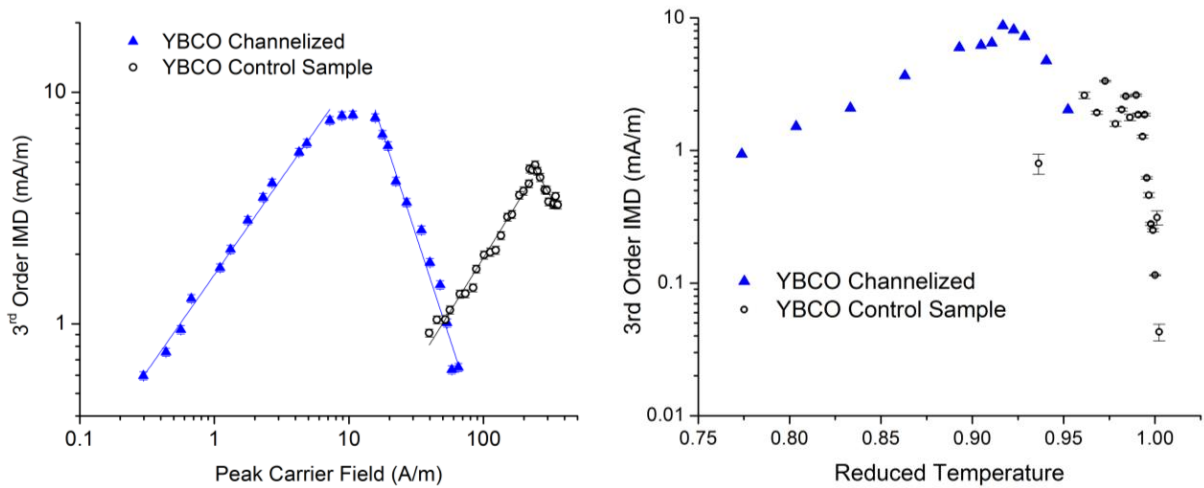

**Figure 4.** 3$^{rd}$ order IMD results for Sample 1 (the 10 T sample) and the control sample at a reduced temperature of T/T$_C$≈0.9. (a) The dependence of IMD (in dBm) on dissipated power (in dBm) below the power level where Q starts to change was fit to a line finding a slope of 0.83±0.02 for the IMD in the channel and 0.94±0.03 for the IMD in the control sample. (b) The temperature dependence of the IMD surface current density in the channel exhibits a NLME peak considerably below $T_C$.

The temperature dependence of the IMD was different for channelized resonators. Typically, there is a peak in nonlinearity just below $T_C$, usually centered at $t=T/T_C \approx 0.97$, which is consistent with the order parameter modulation of the NLME[28]. The NLME IMD peak in Figure 4b occurs at about $t \approx 0.92$ for the channelized resonator. There appears to be two critical temperatures one for the as-grown film and one for the channel about 5 Kelvin lower, a conclusion also reached with previous observations from the Q[26].

When the raster probe was scanned at a constant height across the segment of the transmission line containing the channel for all three samples, an IMD profile along this line revealed the precise location of the channel, along with other centres of nonlinearity such as the high current corners. This is shown later, in Figure 10, where IMD profiles are compared to photoresponse profiles on the same samples. The IMD scan is sensitive to local nonlinearities[29] which can occur around defects and may also reveal hot spots of nonlinearity not associated with defects, such as regions of time reversal symmetry breaking[34].

## 4. Return Loss Photoresponse

By raster scanning a modulated laser beam across the device under test (DUT), the microwave current in a superconducting resonators can be imaged using laser scanning microscopy[35]. In this work, a 638 nm diode laser was deeply amplitude modulated using the reference of a lock-in amplifier (SRS 830 DSP Lock-in Amplifier) which was set to a modulation frequency $f_m$ in the range of 1 KHz to 100 kHz as shown in Figure 5. To understand the purpose of the modulation it is helpful to consider the influence that the laser has on the sample in the absence of modulation.

The DUT is excited with a single microwave loop probe. The inset in Figure 5 shows the $S_{11}$ microwave frequency response of the device with an unmodulated 10 mW laser focused on the channel of Sample 2, and again with the laser turned off. The shift in resonance is due to the photoresponse of the superconductor. The change in RF diode voltage $\delta V$ due to the change in return loss $\delta S_{11}$ is here referred to as the PR signal.

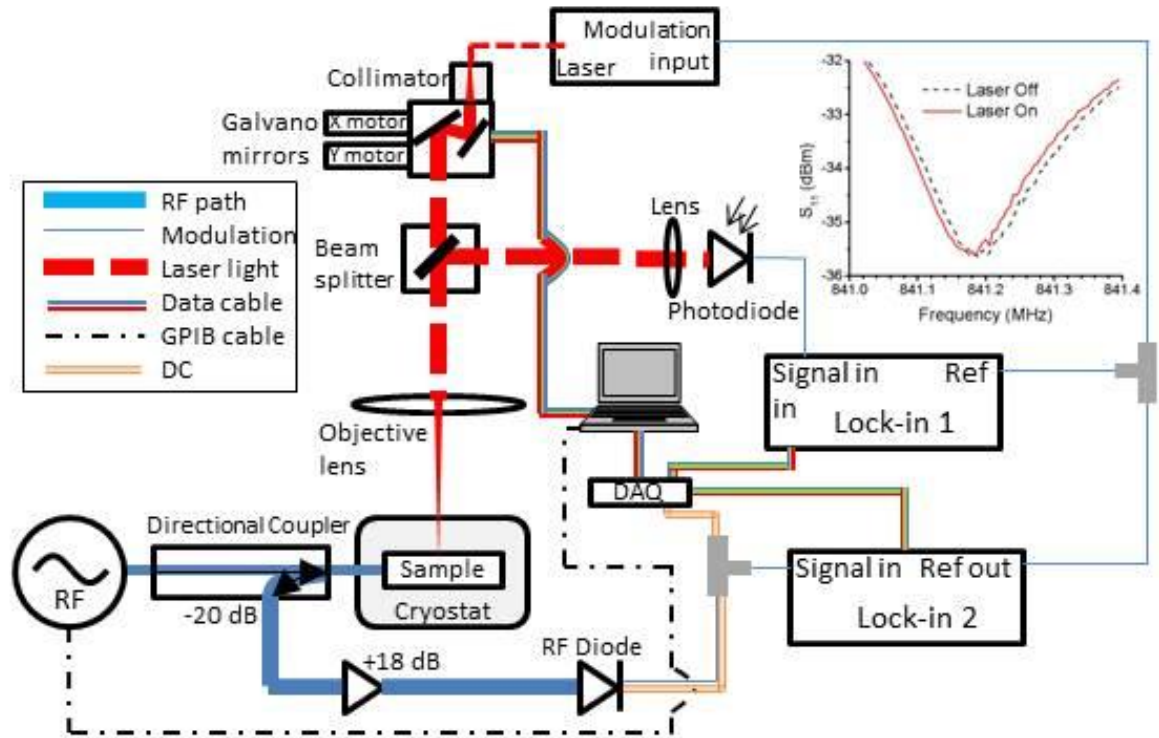

**Figure 5.** (colour on-line) The photoresponse measurement is divided into optical and microwave (or RF) portions. Lock-in 2 provides the modulation tone from its reference, which carries through the laser output causing the sample's microwave response to vibrate at $f_m$. Terminating the reflected microwave signal in the diode leaves the oscillation at $f_m$ intact where it is measured as PR by Lock-in 2. The photodiode receives reflected laser light and its output, also vibrating at $f_m$, is detected by Lock-in 1. The DC output of the diode is proportional to the reflected microwave power and is measured by the computer through the DAQ. The inset shows $S_{11}$ with and without an unmodulated laser. Extremely high microwave power was used to enhance the effect, hence the noticeable asymmetry.

There are three difficulties in measuring $\delta S_{11}$: (1) the shift in $S_{11}$ is small; (2) large amounts of both laser and microwave power are needed to make the shift measurable; and most importantly (3) the response is less localized than the laser beam spot size due to spreading of the heat into the material. These three drawbacks are all addressed by modulating the laser intensity at $f_m$ using the reference source of the lock-in amplifier. Because of the low $f_m$ the superconductor responds bolometrically, modulating the local kinetic inductance at $f_m$[36]. (Bolometric response drops as $1/f_m$, however pair breaking also has weak PR which increases with $f_m$. Most data in this work were taken at $f_m$=20 kHz where the bolometric response still dominates.) $S_{11}(f)$ carries an oscillation at the modulation frequency and is available for measurement by the lock-in amplifier. Though small, this oscillation is above the 2 nV sensitivity of the lock-in amplifier. The sensitivity is sufficient for detection of PR with low laser power, low microwave power, or in parts of the device where the microwave current density is small. Thus, a PR image of the entire DUT is feasible.

PR resolution is limited by the thermal healing length and the laser beam diameter $d_{beam}$ which is determined by the objective lens. The separation between the sample and the objective lens was adjusted so that the beam waist was smallest at the sample. Using information provided in Reference [37], the resolution is a quadrature sum $R = \sqrt{d_{beam}^2 + \Lambda^2}$, where the thermal healing length in μm of the sample is $\Lambda = \sqrt{k/c\rho f_m} \approx 1600/\sqrt{f_m}$, with thermal conductivity $k \approx 10$ W/(m·K), specific heat $c=580$ J/(kg·K), and mass density $\rho = 6{,}570$ kg/m$^3$. Using $d_{beam}=15$ μm and $f_m=20{,}000$ Hz, the PR image resolution is about 19μm. Lower $f_m$ improves sensitivity and degrades resolution since to first order PR increases with the area of the heated region.

Before raster scanning, the microwave frequency dependence of the PR is measured with the laser beam held at a single location. This is done for two reasons: (1) to find the frequency of optimum PR; and (2) to identify the ratio of resistive to inductive contributions to the PR[37] at that location. Figure 6 (a and b) shows PR($f$) with the laser beam above the channel of Sample 1 and (c) in the corner of the control sample. The PR is divided into independent resistive[14,38]

$$PR_R(f) \propto \frac{\partial |S(f)|^2}{\partial \left(\frac{1}{2Q_L}\right)} \cdot \frac{\partial (1/2Q_L)}{\partial P_{laser}}, \quad (1)$$

inductive[14,38]

$$PR_I(f) \propto \frac{\partial |S(f)|^2}{\partial f_o} \cdot \frac{\partial f_o}{\partial P_{laser}}, \quad (2)$$

and dissipative "S-parameter" contrast[38]

$$PR_S(f) \propto \frac{\partial |S(f)|^2}{\partial S(f_o)^2} \cdot \frac{\partial S(f_o)^2}{\partial P_{laser}} \quad (3)$$

components, where $S$ is either $S_{11}$ or $S_{21}$ depending on whether the measurement is transmission ($S_{21}$) or reflection ($S_{11}$)[39]. $f_o$ is the resonant frequency. The first factors in Equations (1) – (3) include the frequency dependence of the S-parameter, whereas the second factors indicate the effect of laser power $P_{laser}$ on the resonator. Resistive and S-parameter PR correspond to resistive nonlinearity. Inductive PR corresponds to a change in kinetic inductance and thus to inductive nonlinearity. In the case of transmission, Zhuravel, *et al.*[38] used $|S_{21}(f)|^2 = \hat{S}_{21}^2 / \left(1 + 4Q_L^2((f/f_o)-1)^2\right)$ to derive expressions for PR($f$) for each component. For a qualitative understanding, these expressions are plotted in Figure 7 using loaded $Q_L=10{,}000$, insertion loss $S_{21}(f_o)=1$, and $f_o=835$ MHz. Only $PR_S$ is non-zero at resonance. In most cases, especially at low RF power, our measured PR goes nearly to zero at resonance indicating that $PR_S$ does not contribute significantly to the photoresponse of these samples, at least at low power, and also providing a convenient way to find $f_o$ (e.g. that frequency where PR is zero).

Attention needs to be paid to the sign of $PR_R$. In microwave reflection, the sign of $\partial |S_{11}|^2 / \partial (1/2Q_L)$ depends on whether the resonator is under- or over-coupled. The following was concluded from simulation using Sonnet 13 (Sonnet Software, Inc., North Syracuse, NY, USA) of the "breathing" of a resonant peak when $Q_u$ is perturbed. When the resonator is over-coupled, $|S_{11}|$ increases (becomes more reflective) as $1/(2Q_L)$ increases, and more power is sent to the RF diode. So the

RF diode voltage is largest at the high point of the laser modulation cycle (when $1/(2Q_L)$ is highest) meaning that the PR is in-phase with the modulation, and therefore $PR_R>0$ for over-coupling. The opposite is true for under-coupling, in which case $PR_R<0$. Since the components of PR add, the symmetry of the PR in Figure 6 indicates the relative significance of inductive and resistive nonlinearity[37]

The resistive PR in Figure 6b is positive in the 75 Kelvin frequency sweeps up to an input power of about +10 dBm, indicating overcoupling up to about +10 dBm. Besides a large difference in magnitude, the shape of PR($f$) over the channel of Sample 1 differs from the control sample. At 75 K, the low microwave power PR of the channel is almost equally resistive and inductive ($PR_R/PR_{total}\approx 0.4$), with the resistive PR shrinking as microwave power increases. At an input power of about +10 dBm the resistive PR crosses through zero and becomes negative for all higher power. The reduction in $PR_R$ with increasing power is in fact an evolution from positive to negative $PR_R$, with $PR_R=0$ at critical coupling. At 82K in Figure 6a the PR of Sample 1 is larger above resonance at all RF powers because so close to $T_C$ the resonator has a low $Q_u$ and is always under-coupled. The PR of the control sample is predominantly inductive with $PR_R$ growing as input power increases.

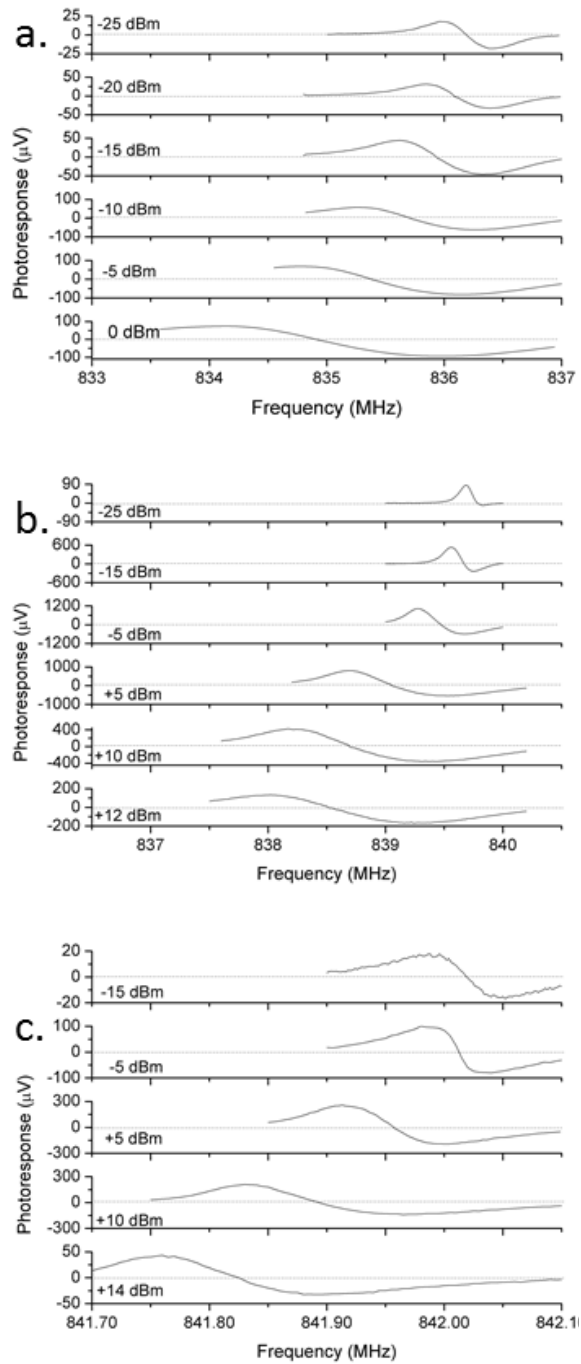

**Figure 6.** Return loss photoresponse frequency scans at (a) 82K and (b) 75K for Sample 1 (10T sample) on the channel at $f_m$=20 kHz; (c) Sample 3 (Control Sample) at the corner at 75K. Comparison of (b) and (c) reveals that whereas the control sample exhibits almost entirely inductive PR, the presence of the channel introduces significant resistive nonlinearity. This distinction is most striking at low power where both measurements were similarly overcoupled on the microwave path.

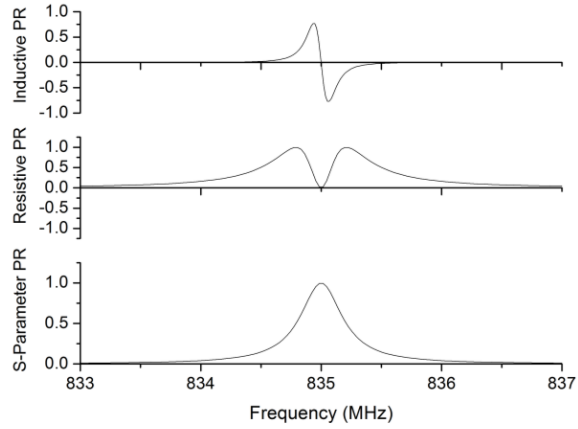

**Figure 7.** Simulated shapes of the frequency dependence of the (top) inductive, (centre) resistive, and (bottom) S-Parameter photoresponse for an over-coupled resonator. $PR_R$ can be positive or negative depending on coupling.

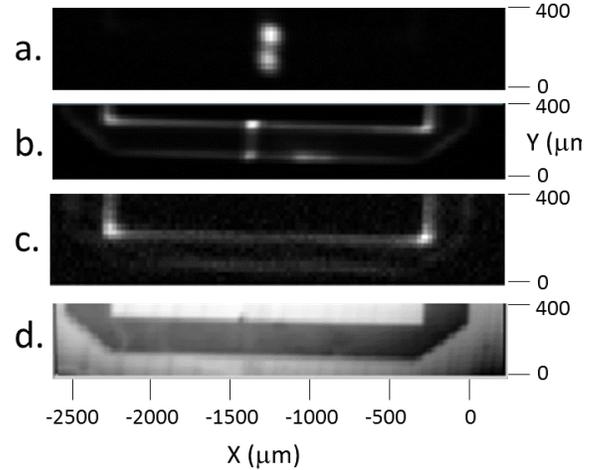

**Figure 8.** Two dimensional raster scans of the photoresponse on the three samples (a) Sample 1, (b) Sample 2, (c) Sample 3 (control sample), and, (d) reflectance image showing the sample, applicable to all three PR images, providing a guide to the PR image locations. Temperature is 75K, RF power is -10 dBm, and $f_m$=20 kHz. Very strong photoresponse is seen in the channel of Sample 1. Some PR occurs in the current crowded corners of Sample 2, which was less irradiated. Each PR image has a different scale.

Using the frequency scan to find the frequency for maximum PR, a fixed-frequency *x-y* scan was then performed. This work focuses on the mid-region of the transmission line, half way along the line between the two ends, where microwave current is high and crowded around the corners. Based on IMD, we expect: the PR in the channel to overwhelm the PR at the corners for Sample 1 and to be similar to the PR at the corners for Sample 2. The PR should be highest in the corners of the control sample. These expectations are consistent with Figure 8. The photoresponse in the as-grown material of the control sample is predominantly inductive at 75 K. $PR_I$ varies as $AJ_{RF}^2\lambda^2\delta\lambda$ where $A$ is the area being heated, and $\delta\lambda$ is the change in penetration depth due to that heating[38]. Assuming $\lambda$ to be uniform everywhere except in the channel, the PR outside the channel occurs where $J_{RF}$ is high. The scans of all three samples clearly show the current crowding that is expected along the edges of microstrip conductors[40], including higher PR along the interior edge than the exterior edge. There may appear to be no PR at the corners of Sample 1, but the sensitivity needed for Figure 8 does not allow the corner PR to be viewed here. In fact, as a point of reference, at 74 Kelvin with a microwave input power of -10 dBm and $f_m$=20 kHz, all three samples exhibit nearly the same PR signal level in the corner.

Microchannels were previously shown to have larger $\lambda$ than the as-grown material[26], leading to the expectation that the channel should have lower $J_{RF}$ due to less tightly crowded current. However, with non-superconducting regions uniformly distributed throughout the channel, the cross-section through which electron pairs flow is reduced. This combination of constricted superconducting cross-section and larger $\lambda$ results in elevated $PR_I$ inside the channel. The enhanced current sensitivity of $R_S$ results in extremely high $PR_R$ in the channel. The reduced $T_C$ of samples with channels is consistent with the reduced pair density in the channel region. Thus a clearer picture of the nature of the channels is revealed by photoresponse.

## 5. Photoresponse and Intermodulation Distortion Together

Both $PR_R$ and $PR_I$ are well understood to depend on the square of the current density ($PR_I \propto J_{RF}^2$, and $PR_R \propto \delta R_S \otimes J_{RF}^2$ as a convolution of $J_{RF}^2$ with $\delta R_S$)[38]. But nonlinearity is directly measured by distortion. Locking onto the modulation frequency transmitted through the IMD should deliver a sharper image and also reveal centres of IMD generation within the sample[22]. The PR carried through the IMD (IMD PR) depends on the variation with laser perturbation of surface resistance $\delta R_S$, $\delta\lambda$, and the nonlinearity scaling current $J_{NL}$. Numerical simulation using the two-fluid model showed that $J_{NL}$ modulation is the strongest contributor to IMD PR[22].

Figure 9 shows a close-up image of the Sample 1 channel in ordinary PR and IMD PR at 82 Kelvin and -12 dBm. The IMD PR was measured in the LSM by driving the sample with two signals at $f_1$ and $f_2$, both in the resonance band, and measuring the IMD at $2f_2-f_1$. The IMD, captured from the analogue output of a spectrum analyser, carries a component at $f_m$, which passes through the RF diode and is measured by the lock-in amplifier. The sharper

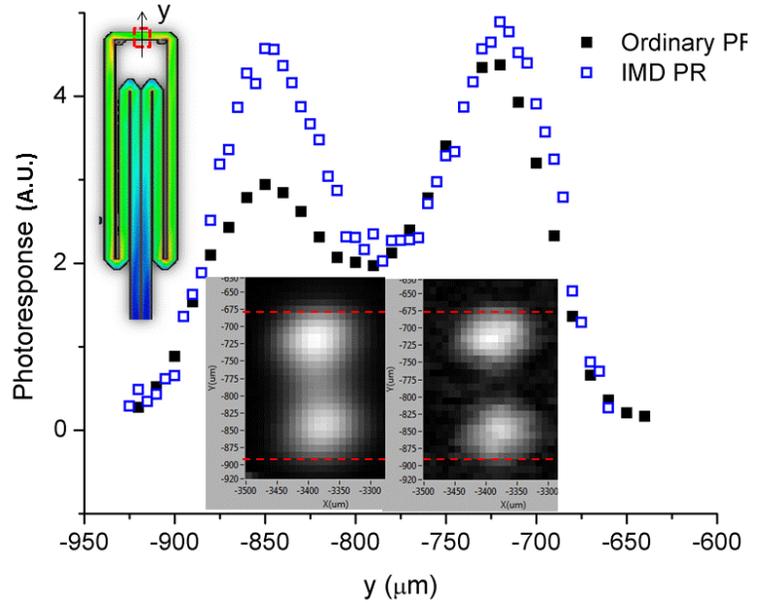

**Figure 9.** (colour on-line) The ordinary photoresponse (*PR*) at 82 K of a single -12 dBm tone (left image) and the photoresponse ($PR_{IMD}$) carried through the IMD (right image) of two -15 dBm tones. The horizontal dashed lines indicate the edges of the transmission line. The dashed box at the top of the inset shows the region of the scans.

IMD PR image in Figure 9 indicates greater sensitivity of $J_{NL}$ to laser perturbation. They also differ in how current crowding is revealed. The IMD PR is not nearly as lopsided from the lower $J_{RF}$ outer edge to the higher $J_{RF}$ inner edge of the transmission line. The IMD PR in the channel is therefore dominated by $J_{NL}$, and by neither $\delta\lambda$ nor $\delta R_S$, which would be scaled by $J_{RF}^2$, resulting in a lopsided curve.

The scanned IMD measured as in Figure 3 can be compared to the PR. Hot spots in IMD are expected to correspond to high $J_{RF}$. The raster probe was moved across each sample producing the local IMD profiles in Figure 10. The width of the IMD peak around the channel indicates the resolution of the loop probe, which has an approximately 400 μm inner diameter. The solid curves in Figure 10 are the ordinary PR, a quantity that is proportional to $J_{RF}^2$.

The channel is by far the strongest source of IMD in Samples 1 and 2. In Sample 1, no IMD is detected at the corners within the available sensitivity, although there is PR there. The different power dependencies of the samples forced the use of different quiescent operating powers for the IMD scans. Sample 1 was tested at -20 dBm because at higher power the resonant peak was distorted. The control sample was tested at 0 dBm, because at -20 dBm its IMD was below the set-up sensitivity (<-133 dBm). Having been exposed to lower ion beam fluence, the Sample 2 channel is less of a nonlinearity hot spot than is the Sample 1 channel. This is evident both in the Sample 2 PR, where the channel PR is less overpowering, and in the IMD, where the IMD was strongest at the channel, though not confined there. Because of the *weaker* channel in Sample 2, both the IMD and the PR could be measured at the corners. One might expect that at higher RF power the IMD scan of Sample 1 will resemble Sample 2. It does not because by -10 dBm the roll-over effect in Figure 4a is well underway in Sample 1.

To the right of the Sample 1 channel, a secondary IMD peak was found which did not correspond to an elevation in PR and was also undetected in the IMD PR scan. PR and IMD are both generated by current density. But the IMD peak to the right of the channel is not associated with any peak in current density detected by PR, which would be the case for a defect. It is however repeatable upon thermal cycles, indicating that this region is susceptible to nonlinear excitation by the 100 kHz probe frequency. Future investigation could employ a static magnetic field to examine Abrikosov fluxon motion in this and other hot spots.

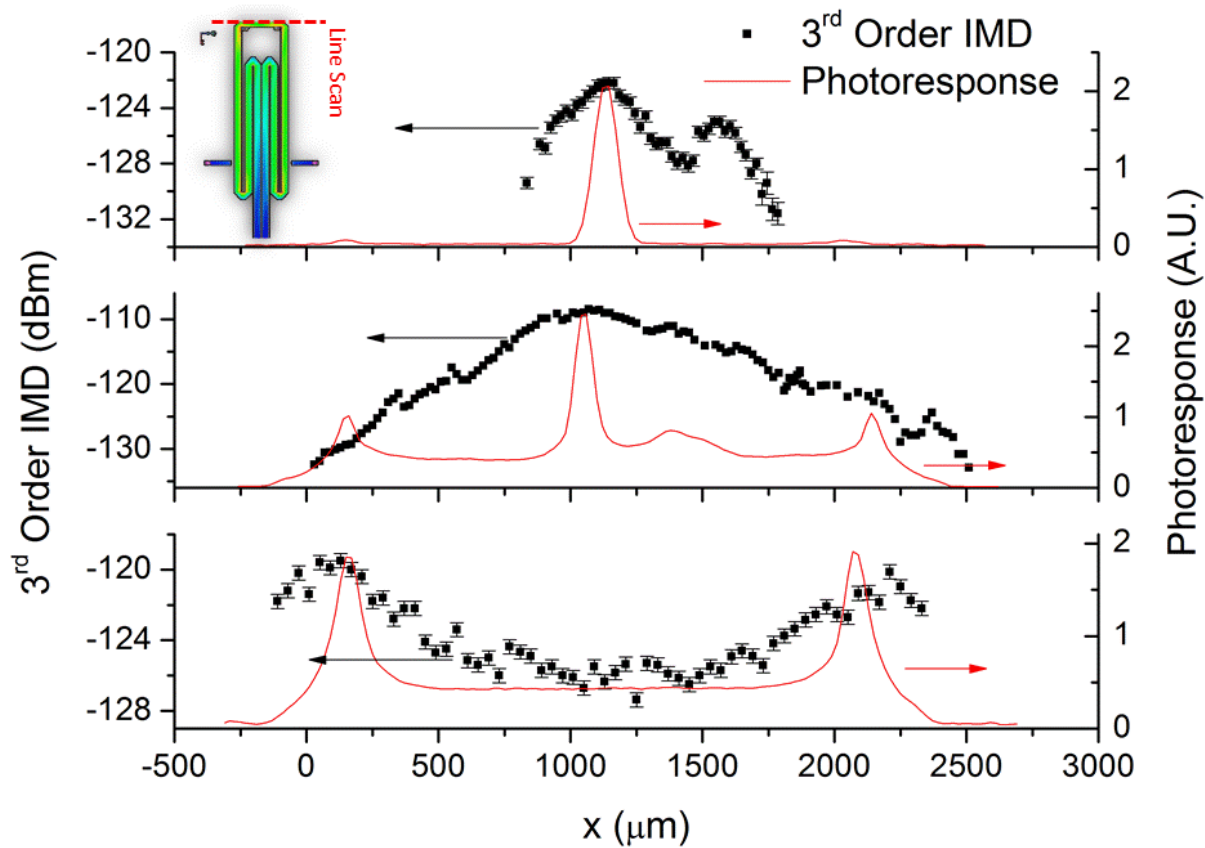

**Figure 10.** (colour on-line) (top) Sample 1; (centre) Sample 2; (bottom) control sample. The image near the top shows the path of the line scan. PR was measured around 75K for each sample. IMD was measured around 80K. Sample 1 was measured at -20 dBm, Sample 2 at -10 dBm, Sample 3 at 0 dBm. The tallest peaks in the top and centre occur at the channel. The outer peaks occur at the corners.

## 6. Conclusions

In this work, progress was made in the measurement, understanding and control of microwave nonlinearity in superconducting resonators. Nonlinearities, which are most often measured globally, were examined here locally both by scanning IMD probe and by LSM. Through scanning IMD, an engineered channel of columnar defects is a hot spot of 3$^{rd}$ order nonlinearity. LSM examination reveals that the nonlinearity is highly resistive compared to the predominantly inductive as-grown film. With a higher density of columnar defects in the microchannel, the superconductor is more photosensitive and the IMD is quantifiably higher at lower microwave excitation power. Despite the highly resistive nonlinearity of the defect channel, at critical coupling there is no resistive photoresponse and the resistive nonlinearity can thus be camouflaged, creating a device whose extreme nonlinearity manifests as purely inductive. In the future, this outcome may serve as a way to realize inductively nonlinear circuit elements.

## 7. Acknowledgements

This work is supported by the NSF-GOALI and OISE Programs through Grant No. ECCS-1158644 and the Center for Nanophysics and Advanced Materials (CNAM). The work at Hope College was supported by Hope College and by NSF Grants DMR-1206149, ARI-0963317, and PHY-1004811. Any opinions, findings, and conclusions or recommendations expressed in this material are those of the authors and do not necessarily reflect the views of the National Science Foundation. The work at Torino was done in the framework of the MESH research agreement between Politecnico di Torino and INFN. Philipp Jung acknowledges the support by the Karlsruhe House of Young Scientists (KHYS) through the KHYS research travel scholarship. Several of the coauthors had extremely valuable in-depth conversations about PR measurement and results with Alexander Zhuravel of the B. Verkin Institute for Low Temperature Physics and Engineering, Kharkov, Ukraine.

## REFERENCES

[1] Eom B H, Day P K, LeDuc H G and Zmuidzinas J, 2012 *Nature Physics* **8,** 623–627

[2] Remillard S K, Yi H R and Abdelmonem A 2003 *IEEE Trans. Appl. Supercond.* **13,** 3797-3804

[3] Ku J, Manucharyan V and Bezryadin A 2010 *Phys. Rev. B* **82,** 134518

[4] Powell J R, Porch A, Kharel A P, Lancaster M J, Humphreys R G, Wellhöfer F and Gough C E 1999 *J. Appl. Phys* **86,** 2137-2145

[5] Ghigo G, Andreone D, Botta D, Chiodoni A, Gerbaldo R, Gozzelino L, Laviano F, Minetti B and Mezzetti E 2005 *Supercond. Sci. and Technol.* **18,** 193-199

[6] Civale L, Marwick A D, Worthington T K, Kirk M A, Thompson J R, Krusin-Elbaum L, Sun Y, Clem J R and Holtzberg F 1991 *Phys. Rev. Lett.* **67,** 648–651

[7] Ghigo G, Laviano F, Gerbaldo R and Gozzelino L 2012 *Supercond. Sci. Technol.* **25,** 115007

[8] Gerbaldo R, Ghigo G, Gozzelino L, Laviano F, Lopardo G, Minetti B, Mezzetti E, Cherubini R and Rovelli A 2008 *J. Appl. Phys.* **104,** 063919

[9] Laviano F, Gerbaldo R, Ghigo G, Gozzelino L, Minetti B, Rovelli A and Mezzetti E 2010 *IEEE Sensors Journal* **10**, 863-868

[10] Oates D E, Nguyen P P, Dresselhaus G, Dresselhaus M S and Chin C C 1993 *IEEE Trans. Appl. Supercond.* **3**, 1114-1118

[11] Remillard S K, Klemptner L J, Hodge J D, Freeman T A, Smith P A and Button T W 1995 SPIE **2559**, 59

[12] Zhang D, Martens S J, Shih C F, Withers R S, Sachtjen S A, Suppan L P, Kotsubo V and Tigges C P (1994), SPIE **2156**, 193

[13] Lu Z, Truman J K, Johansson M E, Zhang D, Shih C F and Liang G C 1995 *Appl. Phys. Lett.* **67**, 712-714

[14] Golosovsky M, Galkin A and Davidov D 1996 *IEEE Trans. Microw. Theory and Techn.* **44**, 1390-1392

[15] Steinhauer D E, Vlahacos C P, Dutta K S, Wellstood F C and Anlage S M 1997 *Appl. Phys. Lett.* **71**, 1736-1739

[16] Lee S –C and Anlage S M 2003 *Appl. Phys. Lett.* **82**, 1893-1895

[17] Mircea D I, Xu H and Anlage S M 2009 *Phys. Rev. B* **80**, 144505

[18] Tai T, Xi X X, Zhuang C G, Mircea D I and Anlage S M 2011 *IEEE Trans. Appl. Supercond.* **21**, 2615-2618

[19] Pease E K, Dober B J and Remillard S K 2010 *Rev. Sci. Instr.* **81,** 024701

[20] Culbertson J C, Newman H S, Strom U, Pond J M, Chrisey D B, Horwitz J S and Wolf S A 1991 *J. Appl. Phys.*, **70** 4995-4999

[21] Newman H S and Culbertson J C 1993 *Micro. And Opt. Technol. Lett.* **6**, 725-728

[22] Zhuravel A P, Ustinov A V, Abraimov D and Anlage S M 2003 *IEEE Trans. Appl. Supercond.* **13,** 340-343

[23] Zhuravel A P, Ghamsari B G, Kurter C, Jung P, Remillard S, Abrahams J, Lukashenko A V, Ustinov A V and Anlage S M 2013 *Phys. Rev. Lett.* **110**, 087002

[24] Laviano F, Gerbaldo R, Ghigo G, Gozzelino L, Minetti B, Mezzetti E 2006 *Appl. Phys. Lett.* **89**, 082514

[25] Laviano F, Botta D, Chiodoni A, Gerbaldo R, Ghigo G, Gozzelino L, Mezzetti E 2003 *Phys. Rev. B* **68**, 014507

[26] Ghigo G, Gerbaldo R, Gozzelino L and Laviano F 2013 *IEEE Tran. Appl. Supercond.* **23**, 1501105

[27] Oates D E, Park S -H, Agassi D, Koren G and Irgmaier K 2005 *IEEE Trans. Appl. Supercond.* **15**, 3589-3595

[28] Lee S -C, Sullivan M, Ruchti G R, Anlage S M, Palmer B S, Maiorov B and Osquiguil E 2005 *Phys. Rev. B* **71**, 014507

[29] Jeries B M, Cratty S R and Remillard S K 2013 *IEEE Trans. Appl. Supercond.* **23**, 9000105

[30] Tai T, "Measuring Electromagnetic Properties of Superconductors in High and Localized RF Magnetic Field," PhD Dissertation, University of Maryland, 2013 (http://hdl.handle.net/1903/14668).

[31] Oates D E, Agassi Y D, Park S -H and Seron D 2006 *J. Phys.:Conf. Ser.* **43**, 556–559

[32] Seron D, Collado C, Mateu J and O'Callaghan J M 2006 *IEEE Trans. Microw. Theory and Techn.* **54** 1154-1160

[33] Sombrin J, Soubercaze-Pun G and Albert I 2013 *Antennas and Propagation (EuCAP), 2013 7th European Conference on,* 25-28

[34] Kaminski A, Rosenkranz S, Fretwell H M, Campuzano J C, Li Z, Raffy H, Cullen W G, You H, Olson C G, Varma C M and Höchst H 2002 *Nature* **416**, 610-613

[35] Zhuravel A P, Sivakov A G, Turutanov OG, Omelyanchouk A N, Anlage S M, Lukashenko A, Abraimov D and Ustinov A V 2006 *Low Temp. Phys.* **32**, 592

[36] Tsindlekht M, Golosovky M, Chayet H, Davidov D and Chocron S 1994 *Appl. Phys. Lett.* **65**, 2875-2877

[37] Zhuravel A P, Anlage S M and Ustinov A V 2006 *Appl. Phys. Lett.* **88**, 212503

[38] Zhuravel A P, Kurter C, Ustinov A V and Anlage S M 2012 *Phys. Rev. B* **85**, 134535

[39] Kurter C, Tassin P, Zhuravel A P, Zhang , Koschny T, Ustinov A V, Soukoulis C M and Anlage S M 2012 *Appl. Phys. Lett.* **100,** 121906

[40] Sheen D M, Ali S M, Oates D E, Withers R S and Kong J A 1991 *IEEE Trans. Appl. Supercond.* **1**, 108-115